# On Image Search in Histopathology

H.R. Tizhoosh[1], Liron Pantanowitz[2]

[1] Department of Artificial Intelligence and Informatics, Mayo Clinic, Rochester, MN, USA
[2] Department of Pathology, School of Medicine, University of Pittsburgh, PA, USA

**Abstract** - Pathology images of histopathology can be acquired from camera-mounted microscopes or whole slide scanners. Utilizing similarity calculations to match patients based on these images holds significant potential in research and clinical contexts. Recent advancements in search technologies allow for implicit quantification of tissue morphology across diverse primary sites, facilitating comparisons and enabling inferences about diagnosis, and potentially prognosis, and predictions for new patients when compared against a curated database of diagnosed and treated cases. In this paper, we comprehensively review the latest developments in image search technologies for histopathology, offering a concise overview tailored for computational pathology researchers seeking effective, fast and efficient image search methods in their work.

## Introduction

The field of "content-based image retrieval" (CBIR) is greater than three decades old [Long2009, Akakin2012]. CBIR is about finding images in a database without metadata by relying on the "content" of the image itself, i.e., the pixels and their spatial relationships, and not by using "metadata" such as keywords and descriptive phrases [Möller2009]. CBIR technology has slowly found its way into computational pathology, especially after the emergence of digital pathology. However, to date there has been no concise overview of the scope of image search and algorithms available for pathology. In this paper, we offer a survey and analysis of different search techniques to thereby help pathologists and other researchers select appropriate CBIR methods to advance their work and innovations.

## Why Image Search?

Visual examination of the "**content**" of tissue samples using a microscope or digital image displayed on a monitor is a task usually relegated to pathologists or computational biologists. Interpretation of such content encompasses some or all of the following steps:

- **Observing Cellular Morphology** (includes cellular structures and their characteristics, architecture based on the arrangement of cells, and background extracellular material)
- **Identifying Cell Types** (subtypes of cells, including normal or abnormal)
- **Noting Cellular Abnormalities (**involving their size, shape, color, and organization)
- **Tissue Integrity** (including preservation of cells & structure, as well as tissue damage or necrosis)
- **Examining Inflammatory Response** (presence of acute and/or chronic host immune reaction)
- **Analyzing Tumor Characteristics** (includes tumor subtype, growth, margins, invasion into adjacent tissues, grade and stage)
- **Using Special Stains and Biomarker analysis** (stains or in-situ hybridization highlight specific cell types, proteins, proliferation index, or other molecules)

It is assumed that "semantic gap" between human experts and computers can be closed if a (properly designed and trained) deep network is employed to extract relevant features in histopathology images [Hare2006, Wan2014, Barz2021]. Hence, one can assume that the aforementioned tasks can be reasonably performed by deep networks if we acquire so-called *deep embeddings* for a tissue image. The semantic gap generally refers to the disparity between low-level image features (such as pixel values, edges, and colors) that deep learning - as well as handcrafted features - extract from unprocessed tissue pixels, and the high-level concepts (such as tissue types and cellular patterns) that pathologists perceive from microscopic images [Schulz2010, Traore2017, Pang2019]. For deep learning to close this gap, it must accurately capture the visual content of tissue images in a way that is interpretable by the pathologist [Zhang2019, Hsu2023]. Learning tissue representation is a major step toward closing this gap [Hemati2021, Sikaroudi2022]. Additionally, employing multimodal domain data, such as pathology reports, can help deep learning attach "context" to its otherwise black-box behaviors when it comes to hierarchical tissue representation in their connectionist topologies [Kalra2019, Ferber2024]

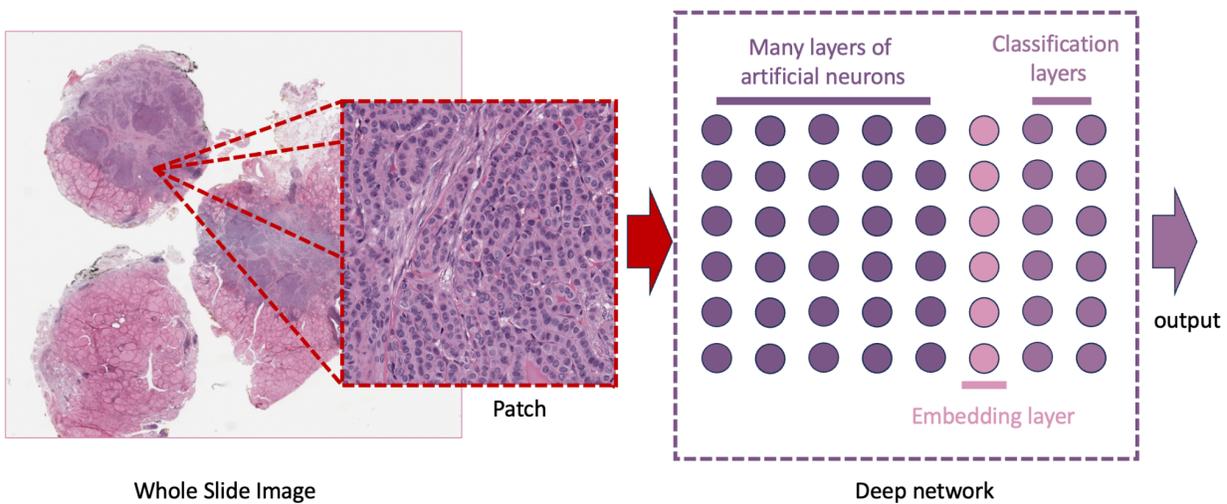

*Figure 1. Small image patches (or tiles) of high-resolution whole slide images are fed into properly trained deep networks to extract a feature vector commonly called "embedding". The output of the network is then neglected. The embedding layer is often positioned before the classification layers at the output of the network.*

## Applications in Histopathology

CBIR has many applications in histopathology, where it can assist pathologists in analyzing and managing large volumes of pathology images to enhance diagnosis, collaboration, education, research, and decision-making processes. In particular, matching images may help perform quality assurance. CBIR can contribute to the diagnosis of diseases by comparing a query image with a database of histopathological images of known cases. By searching and locating visually similar images with known diagnoses, pathologists have more information to identify disease patterns. The same logic applies to treatment planning. Cases with similar diagnostic morphology may provide information to use similar treatment.

CBIR also supports remote consultation and collaboration between pathologists. Retrieving similar cases for/in a remote database, a new tool in telepathology, pathologists can discuss and share knowledge, enabling second opinions and enhancing consultation and/or consensus to resolve challenging cases. Moreover, finding and analyzing similar tissue patterns can be a teaching tool in histopathology education, because it allows trainees to study and compare various histological patterns and diseases, sometimes even using rare, archived cases.

Image search is a crucial tool in research, allowing the exploration of large histopathological image archives to discover and identify novel patterns, trends, and correlations among diseases and their histological characteristics. By leveraging this approach new associations, biomarkers, and predictive factors can be discovered in collaboration with other data mining technologies.

## Divide and Conquer: How AI can Digest Whole Slide Images

In computer science, "Divide and Conquer" (**D&C**) (see Figure 1) is a general problem-solving idea that proposes to break down a complex problem into smaller, more manageable subproblems that can be solved independently; combining these individual solutions in order to solve the original "unsolvable" problem. Whenever we talk about a solution for complex problems, a reasonable computation time is implied. However, nobody can wait a prohibitively long time for a computer to calculate the solution for a difficult problem.

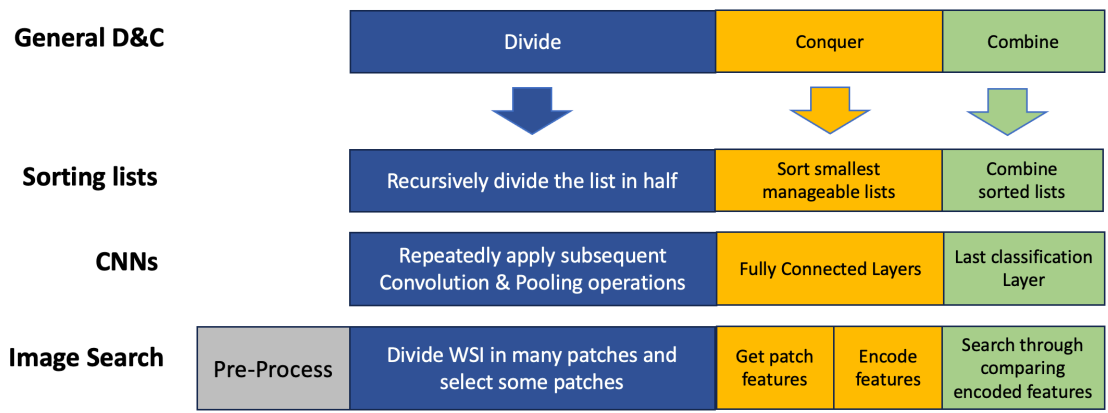

*Figure 1. Divide & Conquer (*D&C*). Sorting a long list of numbers, the architecture of convolutional neural networks (CNNs), and stages of comparing whole slide images (WSIs) represent different examples for applying D&C when we are dealing with an extremely difficult problem. Although the general D&C concept does not involve any pre-processing of data, in image search we often add pre-processing (e.g., tissue segmentation).*

**Example: Sorting Numbers -** If we have one number, no sorting is necessary. In fact, every single isolated number is sorted by itself. However, what happens if we get two numbers? Well, we can keep them as they are if they appear in the right order, or rearrange them if they do not. Sorting small lists of numbers is not a difficult problem. But what happens if the list of numbers is extremely long and contains millions of numbers? In this case we can begin by swapping numbers by looking at the adjacent pair of numbers. Even after reaching the end of the list and having undertaken millions of swaps, the list may still not be fully sorted. Figure 2 illustrates a simple example for sorting.

- **Divide**: The list is recursively (i.e., repeatedly in a self-similar manner) divided into smaller sublists. This step involves breaking down the problem into manageable parts, until the subproblems become simple enough to be solved directly (i.e., this yields a list with only few numbers).

- **Conquer**: Each sublist is sorted independently. This typically involves applying the same D&C technique recursively to the sublists until they can be easily sorted or when reaching a "base case" where a straightforward solution is obvious (e.g., one single number that is automatically sorted).
- **Combine**: The sorted sublists are combined to obtain a final list where all numbers appear in the right order. This step involves merging sublists to assemble the sorted list, (i.e., the solution).

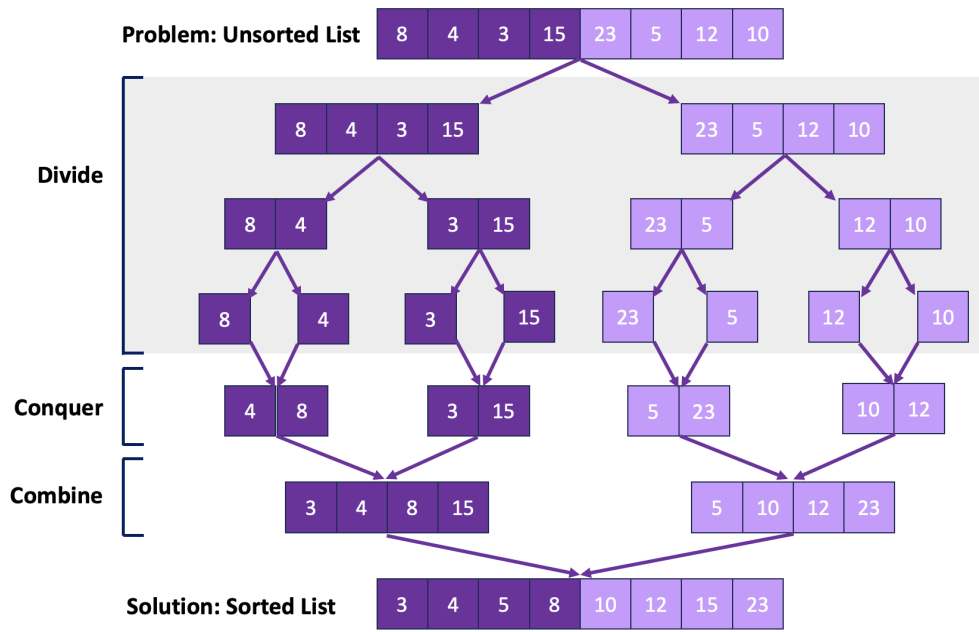

*Figure 2. Sorting a list of unsorted numbers through D&C paradigm. In complex problems, conquering the problem to generate a sorted list is not possible without a "Divide" stage.*

For one million numbers, a method that does not use D&C and sorts the list by swapping adjacent pairs would take approximately "one million by one million = one billion" swaps. This would take a prohibitively long waiting time for the user. Instead, if a method that does use D&C would take approximately "one million by logarithm of (one million) = six million" swaps. Now, instead of one million numbers, imagine we need to classify a whole slide image of size 50,000 by 50,000 pixels. This is equivalent to 3 x 50,000 x 50,000 = 7.5 billion pixels. Furthermore, when analyzing such a gigapixel file we are not just sorting numbers, but we have to perform much difficult tasks like recognizing relationships between adjacent numbers in all directions for a plurality of variations of different tissue types in all organs. In computer science, we call this a "NP-Hard" problem; jargon that signifies the problem cannot be (easily) solved via conventional approaches.

Convolutional neural networks (CNNs), which enabled the revival and tremendous success of AI, are an excellent implementation of the D&C concept. Successive layers of convolutions (filtering) and pooling (breaking down the image into smaller images) make the problem (image identification) manageable for a simple network of a few fully connected layers (the conquer part) such that a final layer can then combine everything into a decision. Searching for WSI content needs its own Divide and its own Conquer to be both fast and memory efficient.

# The "Divide" in Histopathology Image Search

Presently, even with multicore computers, ample memory, and clusters of powerful GPUs, we still face serious limitations when it comes to processing large WSI files for any purpose - whether using AI or not. Consequently, current computers are unable to instantaneously analyze a WSI in its entirety, determine the contents based solely on pixel data, and accordingly provide a diagnosis or prognosis. As for image search and many other tasks, a practical approach involves dividing WSI files into multiple "patches" (smaller sub-images) and processing them individually. However, due to the gigapixel scale of WSIs, patching schemes (which embody the 'divide' operation) may generate numerous patches of any given size, typically ranging from 224 by 224 pixels up to 1024 by 1024 pixels. Consequently, GPU processing of these patches becomes time-consuming and expensive.

The "Divide" of WSIs can be performed in two ways: 1) sub-setting, or 2) patching (see Figure 3). Sub-setting approaches attempt to find one large portion of a WSI that contains the region or abnormality of interest [Barker 2016]. This approach, while potentially effective in reducing the processing of excessive patches, lacks popularity due to its requirement for supervision. Like other supervised methods, it restricts the "Divide" process to the specific primary site and diagnosis for which it has been trained. In addition, due to the large number of tissues and heterogeneity of diseases, and possible presence of multiple concomitant abnormalities, sub-setting may fail in many scenarios. Hence, most papers that have implemented any notion of "Divide" for any type of WSI processing have employed patching.

Table 1 – Comparison of sub-setting and patching for dividing WSIS.

|  | **Divide via Sub-Setting** | **Divide via Patching** |
|---|---|---|
| Learning | Supervised | Unsupervised |
| Primary site | Needs customization | Agnostic |
| Abnormality inclusion | One | Multiple |
| Risk of missing a second abnormality | Very high | Low |
| Risk of missing a small abnormality | Very high | Low |
| Works best for… | One large tissue sample containing one connected segment of abnormality if you have sufficient delineated cases | Any tissue sample as long as the abnormality is not too small |

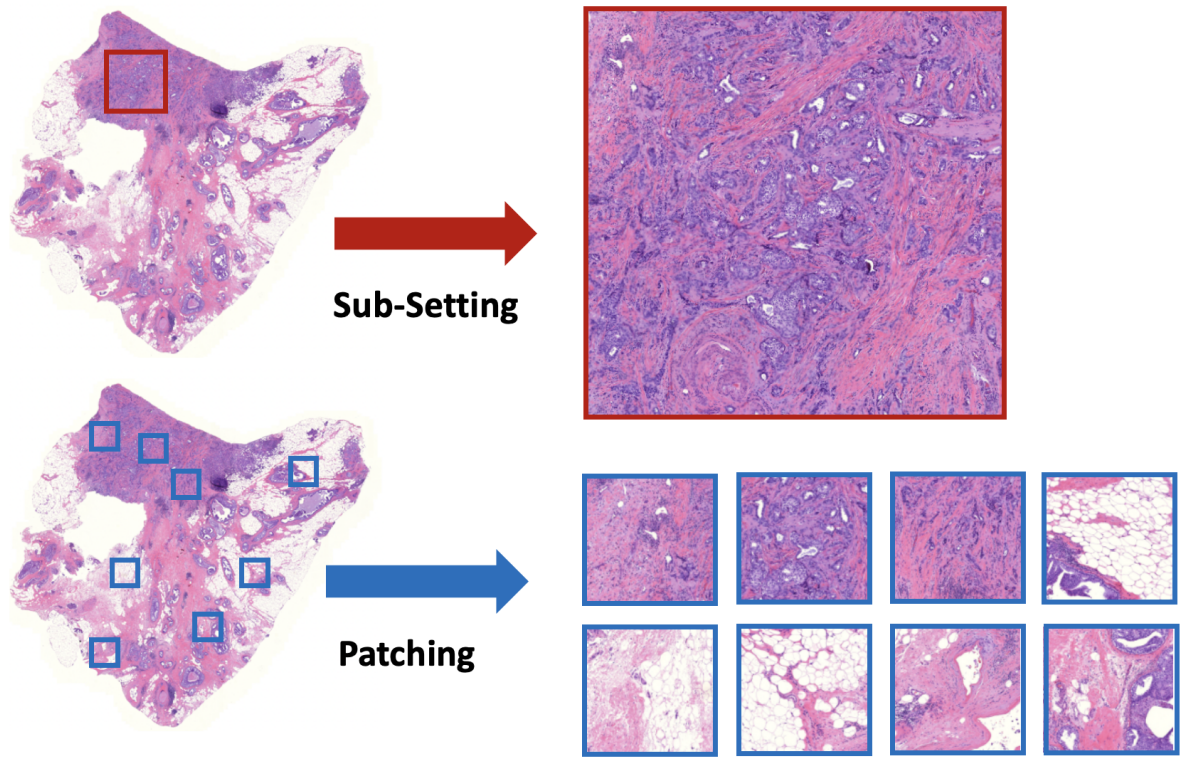

*Figure 3. The "Divide" of a WSI can be performed either as a sub-setting problem (top) or a patching problem (bottom). [Image source: TCGA, Case UUID b8a44fdf-9cb9-4123-9ab0-4bc198921fee, breast invasive carcinoma, file: TCGA-OL-A5RU-01Z-00-DX1.A48CAF2D-9310-4611-B27D-400F3A324607.svs].*

The "*Divide*" of a WSI, i.e., the patching procedure for search purposes, must satisfy the following requirements to be practical in digital pathology:

1. **Universality**: It must be unsupervised (a Divide approach subject to training will limit usability)
2. **Tissue Size/Shape Independence**: It must process the image of all types of specimens, including core biopsies and excisions (different shapes and sizes of tissue should be handled automatically)
3. **Diagnostic Inclusion**: It should not miss any diagnostically relevant tissue part (even small but unique patches should be extracted for indexing)
4. **High Speed**: It should extract patches in a timely manner (brute force methods based on expensive operations like deep feature extraction will not be feasible)
5. **Storage Efficiency**: It should aim to extract a minimal number of patches. Additionally, the encoding process should not demand excessive storage (the necessity for low storage of deep embeddings and any encoding method is crucial in the context of digital pathology)

Any new image search strategy must put forward a new patching paradigm that ideally satisfies all of these aforementioned conditions.

## "Conquering" the Patches

The "Conquer" part of image search in histopathology has mainly two parts: 1) to get feature vectors for each patch, and 2) encode the feature vector for efficient storage and processing. Using deep embeddings as features for digital images has been under intensive investigation in recent years [Shen2017, Garcia2018, Bidgoli2022, Tommasino2023]. It is generally assumed that deep learning has closed the semantic gap between subjective image assessment and computerized processing of digital images [Qayyum2017, Deorukhkar2022]. Training a suitable topology from scratch, fine-tuning a pre-trained network, or simply using zero-shot features from a backbone network are all possible venues to conquer the patches of a WSI. As we will see, different search strategies have explored all of these possibilities. However, conquering WSI patches that come from the Divide solely through deep feature extraction is not possible for practical deployment. Going fully digital, a natural prerequisite for image search, requires high-performance storage that is a considerable investment for many clinics and hospitals [Hanna2022, Eccher2023].

Encoding is another indispensable part of conquering WSIs. Here we understand encoding rather in its original information-theoretical meaning related to coding theory involving data compression. We cannot simply save all of the feature vectors of all patches of all WSIs. To store a vector of say 1024 real-valued numbers between -10 and +10, we would need 1024 numbers * 4 bytes/number = 4096 bytes if we use single-precision floating-point numbers. For double precision the storage of one vector increases to 8192 bytes. If we manage to extract 100 representative patches on average from each WSI, we would need at least 400 kilobytes to "index" the WSI. In contrast, a binary feature vector would need 100*1024 elements * 1 bit/element = 100024 bits = 12 kilobytes. Hence, binary encoding would reduce storage by 97%. The higher processing speed and diminished GPU demand are other advantages of encoding.

## Archive Size

It has been assumed, for unknown reasons, that digitized histopathology archives to be searched are very large. Whereas the number of biopsy cases per year, and average number of glass slides per patient, to be scanned in a typical pathology laboratory do indeed lead to emergence of large archives, we do not have to search these large archives (and as of 2023, large digital histopathology archives are rather the exception). It can very well be that designing and validating matching algorithms in small, but well curated collections is much more feasible and beneficial. As large heterogeneous pathology archives remain to be widely established, it may not make sense to talk about them in abstraction.

## Performance Metrics

Evaluating the performance of image search algorithms is more challenging than image classification algorithms. As a classification model provides specific and distinct output as a decision, its evaluation in terms of accuracy is rather straightforward (i.e., true and false). Although the outputs of image search may be looked at as true/false as well (e.g., in terms of primary diagnosis), search cannot be comprehensively evaluated like classification as it provides a set of results (i.e., search results). In the literature, several performance metrics[1] are used to evaluate the effectiveness of CBIR systems [Buttcher2016, Zhou2018, Müller2001]. Below is the list of most used performance measures. As per literature, precision and recall, and their harmonic average, F1 score, are de facto standard for most retrieval tasks as they build on sensitivity and specificity [Buttcher2016] (see Figure 4).

## Performance per Class/Subtype

- **F1 Score** is the harmonic mean of precision (how many of the retrieved images are actually relevant to the query, i.e., low false positive rate) and recall (how many of the relevant images were actually retrieved, i.e., low false negative rate), providing a balanced measure of both metrics. For detailed analysis, one may provide the values of precision and recall as well.
- **Precision-recall curve** plots the precision values against different recall levels. A high area under the curve (AUC) represents both high recall and high precision.
- **P@K and R@K** may also be reported to measure the precision and recall for the top K retrieved images, respectively.
- **Majority@K** measures the accuracy of image search for the top K search results. In contrast to the "top K" accuracy in computer vision (search function would be successful if at least one of the retrieved images was relevant/correct). The majority@K considers the search successful only if at least 51% of the retrieved images were relevant/correct. This measure was first used to evaluate search results in histopathology in the Yottixel paper [Kalra2020a].
- **Relevance feedback/judgements**, in contrast to all other previous metrics and measures, evaluates the performance of image retrieval based on user feedback, i.e., the pathologists' judgements. Image retrieval based on relevance feedback has shown considerable performance increase [Zhou2003].
- **Mean Opinion Score (MOS)** is "a numerical measure of the human-judged overall quality of an event or experience" and is extensively used in the telecommunication industry [Huynh2010]. It was first used to evaluate the correlation between pathologists' feedback and search results [Kalra2020a].

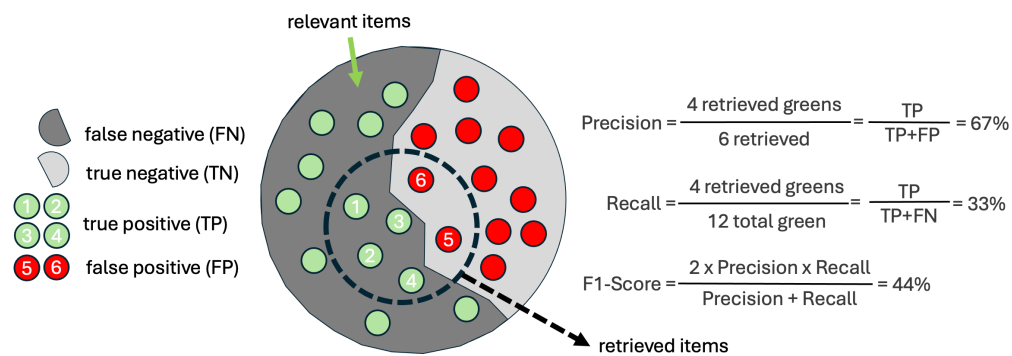

*Figure 4. Precision and recall (derive from true/false positives and true/false negatives) are the base for precision (consistency) and recall (accuracy). F1-score, as the harmonic average of precision and recall is an aggregated measure of both.*

---

[1] A metric space, mathematically speaking, is a set together with a notion of distance between its elements. The distance is measured by a function called a metric or distance function. A metric space has some mathematical properties [Chen2009] that may not be satisfied by all evaluation methods discussed here. Hence, we use the word *metric* in a generic, less strict sense.

## Overall Performance

- **Mean Average Precision (mAP)** calculates the average precision at different recall levels and then takes the mean of those average precisions. mAP provides a measure of the overall effectiveness of the search.
- **Mean Average Precision at K (mAP@K)** focuses on the precision at a specific value of K (e.g., mAP@5). It measures the average precision of the top K retrieved images.
- **Normalized Discounted Cumulative Gain (NDCG)** evaluates the ranking quality of retrieved images. It is often used to measure performance of web search engine algorithms. NDCG considers both relevance and rank position, giving higher weight to relevant images that are ranked higher.
- **Macro and micro averaging** of any metric are used mainly for multi-class classification. Macro averaging treats the contributions of all classes equally to the final averaged metric. Micro averaging treats the contributions of all samples equally to the final averaged metric. If you have an imbalanced dataset (which is often the case in histopathology), micro averaging generally provides higher values than macro averaging. It is recommended to avoid micro averaging for imbalance datasets [Narasimhan2016].
- **Average normalized modified retrieval rank (ANMRR)** is a normalized ranking method. The NMRR score ranges from 0 to 1, where 0 indicates perfect retrieval.

When searching histopathology digital images, the most conservative performance evaluations are majority@K. Relevance feedback from user, i.e., pathologist, is obviously crucial but as image search in histopathology does not seem to have been implemented for clinical utility, there are no user studies or statistics available in this regard. The F1 score stands out as a more comprehensive measure for search to summarize majority@K because it effectively quantifies both precision (minimizing false positives) and recall (minimizing false negatives) and is based on parametrized F measure [Hersh2020]. Whereas majority@K goes beyond classification and top-k metrics, and rigorously requires the majority of retrieved images to be correct/relevant (Figure 5), the relevance feedback is the ultimate evaluation representing practical implementation of the Turing Test.

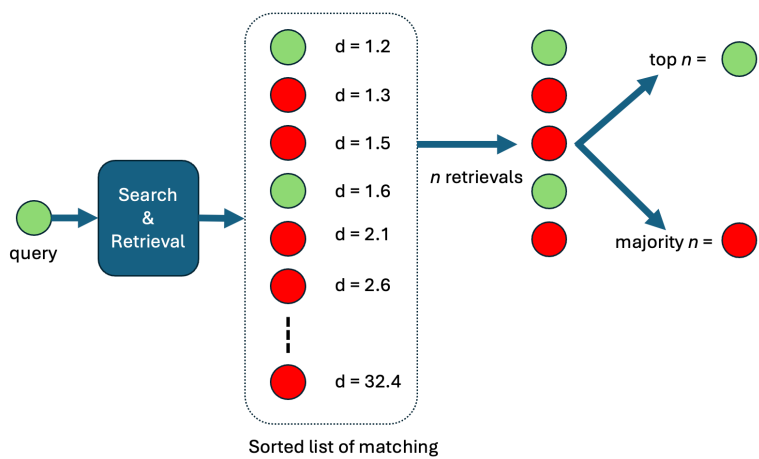

*Figure 5. A query of certain class (here green circles) is matched to all instances in an archive by distance calculation. The distance vales are then sorted in ascending order (most similar at the top of the sorted list). The n similar cases at the top of the list will be retrieved as the output. Then, the "top n" accuracy as common in computer vision will assign 'green' because at least one retrieved circle, among 5, is green. However, the 'majority n' scheme assigns 'red' to the query because 3 out of 5 retrievals are red.*

# Existing Image Search Engines

There are many search techniques that have been applied to histopathology images [Sharma2012, Sridhar2015, Zheng2018]. Based on our background review, six of them may be of relevance for today's needs. We look at hashing-based image retrieval as described in [Zhang2014] (we call it short HBIR), visual dictionary (or bag of visual words, BoVW [Zhu2018]), Similar Medical Images Like Yours, SMILY [Hegde2019], Yottixel [Kalra2020a], Self-supervised Image Search for Histology, SISH [Chen2022], and Retrieval with Clustering-guided Contrastive Learning, RetCCL [Wang2023]. These are the most recent and most relevant search approaches to histopathology images regarding the current needs and state of technology.

*Architectural Design of Search Methods*

Based on previous considerations we look at five stages of image search pre-processing, dividing the WSI, conquering the search task (consisting of a network for feature extraction, and encoding), matching, and optional post-processing. Table 1 provides an overview of the selected search schemes and their structures.

**PreProcessing -** The most common preprocessing is segmentation of tissue region. Not having segmentation is an indication that there is no Divide. HBIR and SMILY do not have segmentation, hence do not offer any Divide. Such methods are based on manual or exhaustive patch selection and are generally inefficient.
**Divide -** Patching is the pivotal stage for image search. Without patching there is no proper indexing and hence no search. HBIR and SMILY, as mentioned, offer no Divide. BoVW samples visual words in both deterministic and stochastic ways. However, the literature has ignored the potentials of BoVW for histopathology and we could not find many papers that employed a visual dictionary. Yottixel is the only method that offers a new Divide, unsupervised, to get a "mosaic" of patches. SISH borrows the Yottixel's mosaic. RetCCL also uses Yottixel's mosaic, but instead of employing a color histogram for clustering it uses deep features, an approach that makes the indexing very slow.
**Conquer (Network) -** HBIR did not use deep features at all. BoVW may use any type of features. SMILY and RetCCL used custom-trained networks. SISH uses two networks, a DenseNet like Yottixel and a custom-trained autoencoder, but performs poorly. Yottixel uses DenseNet. Assuming that one custom-trained network will be suitable for analyzing an entire histopathology database appears to be a rather restrictive factor and not an advantage.
**Conquer (Encoding) –** This entails additional measures to make processing and matching of features more effective and/or efficient. HBIR uses hashing, a well-established set of algorithms to convert real-valued numbers into a series of zeros and ones. Visual dictionaries use counting of visual words as a type of encoding to create a compact histogram. Yottixel introduced the novel concept of "barcoding" feature vectors through one-dimensional derivatives (i.e., encoding the changes of deep features) [Tizhoosh2015, Kumar2018]. SISH also uses Yottixel's barcoding although it does not cite the original works. RetCCL does not use any encoding.
**Combine -** In terms of search and matching, the Combine approach is favored for match.
**PostProcessing -** HBIR, BoVW, SMILY and Yottixel do not use any post-processing. SISH used a multi-stage ranking approach to rearrange search results. RetCCL used the same ranking algorithm as SISH. Although any domain knowledge or unsupervised method may be used to improve search results, ranking search results (which have been already ranked by the search engine) appears to be ill-motivated. The primary concern in questioning the value of ranking, especially in the context of SISH's utilization, as a post-processing task in histopathology is that it eliminates the possibility of comprehensive WSI-to-WSI matching. Matching patients becomes challenging, as single patches in two WSIs may exhibit similarities or dissimilarities. To enable thorough tissue comparison between two patients, extensive comparisons of all representative patches in the respective WSIs are necessary. The SISH ranking reduces search to classification. After search, all patches are again processed and ranked outside the search engine.

Table 2 - Structure of the most recent strategies for searching digital histopathology archives. Not having tissue segmentation leads to lack of Divide, a major drawback of HBIR and SMILY (black cells). Not using encoding is a disadvantage as well (gray cells). Both SISH and RetCCL use Yottixel's mosaic for Divide. SISH uses the complete Yottixel processing chain and adds an autoencoder and a tree (yellow cells). Ranking search results appears to compensate for inferior search results but adds computation overhead, eliminates the possibility of WSI-to-WSI matching, and reduces search to classification (red cells).

|  | Pre-Proc. | Divide | Conquer Network | Encoding | Combine | Post-Proc. |
|---|---|---|---|---|---|---|
| **HBIR** | ■ | ■ | None | Hashing | Hashcode matching | None |
| **BoVW** | Segment | Visual word sampling | Any feature extraction method | Counting | Histogram matching | None |
| **SMILY** | ■ | ■ | Custom | None | Feature matching | None |
| **Yottixel** | Segment | Mosaic | DenseNet | Barcoding | Barcode matching | None |
| **SISH** | Segment | Yottixel's mosaic | DanseNet | Yottixel's barcoding | Barcode matching | Ranking |
|  |  |  | Autoencoder |  | Tree matching |  |
| **RetCCL** | Segment | Yottixel's mosaic | Custom | None | Feature matching | Ranking |

*Strengths and Weaknesses of Search Methods*

The strengths and weaknesses of the aforementioned six search strategies were analyzed (Table 3). As Yottixel is the only complete search engine with its own Divide & Conquer, most strengths are assigned to it. HBIR and SMILY do not have any Divide (they cannot index WSIs), and both SISH and RetCCL use Yottixel's. HBIR is certainly superior to SMILY for patch search as it uses hashing (binarization of features). SISH is a re-implementation of Yottixel as it borrows the entire Yottixel chain, mainly both Divide (mosaic) and encoding in Conquer (barcoding via MinMax algorithm) [Sikaroudi2023]. RetCCL appears to only propose a trained network and use simple matching. RetCCL, like SISH, uses Yottixel's mosaic but it makes it very expensive by replacing color histograms with deep features.

Table 3 - Strengths and weaknesses of the search schemes.

| Search scheme | Strength | Weakness |
|---|---|---|
| **HBIR** | • Hashing (fast processing) | • No divide (no WSI processing)<br>• No deep features |
| **BoVW** | • No patching, no sub-setting<br>• Unsupervised | • Requires custom setting (size of visual words, size of dictionary)<br>• Requires training of custom topology |
| **SMILY** | • Custom network training | • No divide (no WSI processing)<br>• No encoding |
| **Yottixel** | • Novel divide<br>• Novel binary encoding<br>• Patch and WSI matching<br>• Validated with majority@k<br>• Validated by three pathologists | • Mosaic needs configuration (number of clusters and sampling rate)<br>• Requires network selection<br>• Commercially patented [USPatent2020] |
| **SISH** | • Using a tree for implementation | • No new Divide (uses Yottixel's mosaic)<br>• Autoencoder does not contribute to accuracy<br>• No thorough WSI matching<br>• Tree indexing needs a lot of storage<br>• Search perhaps slow due to integer indexing<br>• Commercially patented [USPatent2020] (due to using Yottixel's barcoding) |
| **RetCCL** | • Custom network training | • No Divide (uses Yottixel's divide)<br>• Expensive Divide (uses deep features instead of color histogram)<br>• No encoding |

*Validation of Search Methods*

Yottixel, SISH and RetCCL use TCGA (The Cancer Genome Atlas) for training. Yottixel has not used any of TCGA for training and employs the entire TCGA database (both diagnostic slides and frozen section) for validation [Kalra2020b]. As such, the validation of Yottixel is the most trustable method. SISH and RetCCL use part of TCGA (diagnostic slides only) both for training and testing. We could not verify the data hygiene for SISH and RetCCL (i.e., strict separation between training/testing data).

Lahr et al. have conducted a largescale analysis and validation of BoVW, Yottixel, SISH and RetCCL using internal and external data [Lahr2024]. They report that BoVW and Yottixel offer advanced search solutions with a blend of high speed and efficient storage, presenting valuable capabilities. However, achieving enhanced accuracy requires integration with a well-trained backbone network and adjustments to the primary site. Yottixel, being a commercial product, allows researchers more freedom to explore various BoVW variants. Manual settings in Yottixel's mosaic, particularly regarding cluster number and sampling percentage, could benefit from automation. A logarithmic barcode comparison approach could further boost the speed of median-of-minimum Hamming distance calculations in Yottixel. Lahr et al. also conclude that SISH, as a Yottixel variant, departs from Occam's Razor principles, introducing speed and scalability challenges due to unnecessary complexity. SISH's reliance on vEB trees with exponential space requirements renders it impractical for large datasets, hindering loading and processing of terabytes of data. RetCCL, though labeled a search engine, primarily focuses on the CCL network, lacking expressive embeddings for tissue morphology and, therefore, struggling to qualify as an effective search engine. Lahr et al. provide a ranking of some variations of these search methods (see Table).

Table 4 – Lahr et al. ranked the performance of different search engines based on their accuracy (F1-score for top-1, majority of top-3, and majority of top-5), indexing time, searching time, failures, and storage (performance ranking between 1 (best) and 6 (worst)).

|  | Top-1 | MV@3 | MV@5 | Indexing time | Searching time | Failures | Storage | Total Ranking |
|---|---|---|---|---|---|---|---|---|
| Yottixel* | 3 | 2 | 1 | 2 | 1 | 1 | 2 | **1.71** |
| Yottixel-KR[1] | 2 | 1 | 2 | 2 | 1 | 2 | 2 | **1.71** |
| Yottixel-K[2] | 1 | 3 | 4 | 2 | 1 | 2 | 2 | **2.14** |
| BoVW | 6 | 6 | 6 | 1 | 2 | 2 | 1 | **3.43** |
| SISH* | 4 | 4 | 3 | 3 | 3 | 4 | 4 | **3.57** |
| SISH-N[3] | 5 | 5 | 5 | 3 | 3 | 4 | 4 | **4.14** |
| RetCCL-N[3] | 8 | 8 | 7 | 4 | 4 | 3 | 3 | **5.28** |
| RetCCL* | 7 | 7 | 8 | 4 | 4 | 3 | 3 | **5.43** |

* As originally proposed
[1] DenseNet replaced by KimiaNet, and using ranking after search
[2] DenseNet replaced by KimiaNet
[3] SISH with no ranking after search
[4] RetCCL with no ranking after search

## Multimodal Search in Histopathology

Multimodal information retrieval is not as old as image retrieval but has been investigated extensively [Hauptmann1997, Srihari2000]. Combining images with other modalities is less mature [Chang2003, Kumar2013]. Indeed, studying multimodal retrieval in conjunction with medical images is a newer field [Cao2014, Kitanovski2017]. Only a few works have gone beyond image search and integrated other modalities into search. For instance, textual metadata has been combined with histopathology images in a cross-modal learning framework [Maleki2022]. Also, molecular data such as RNA sequences have shown promise to increase search accuracy when combined with WSIs [Alsaafin2023]. It is hard to argue for the usefulness of image search if WSIs, patches, or microscopic snapshots are only accompanied with a primary diagnosis. The value of image search degrades into a fancy classifier that cannot even match the accuracy of a classifier. The real value of CBIR arises when combining tissue images with diagnostic reports, clinical notes, patient demographics, laboratory test values, molecular data, radiology images, and patient outcomes. As of today, there is no multimodal search for pathology.

## Image Search and Foundation Models

The domain of search and retrieval may initially seem distinct from artificial neural networks. However, one may argue that this distinction is only superficial. Traditional search engines typically function through lookup tables, which are generated using specific functions and algorithms to locate desired information. This implies that while we partially construct these tables, but we do explicitly utilize tables to locate information. On the contrary, neural networks—whether shallow or deep—can be likened to underline{soft tables}. Within these networks, the network structure itself or intricate pathways within it assist in locating and delivering the desired information (see Figure 6).

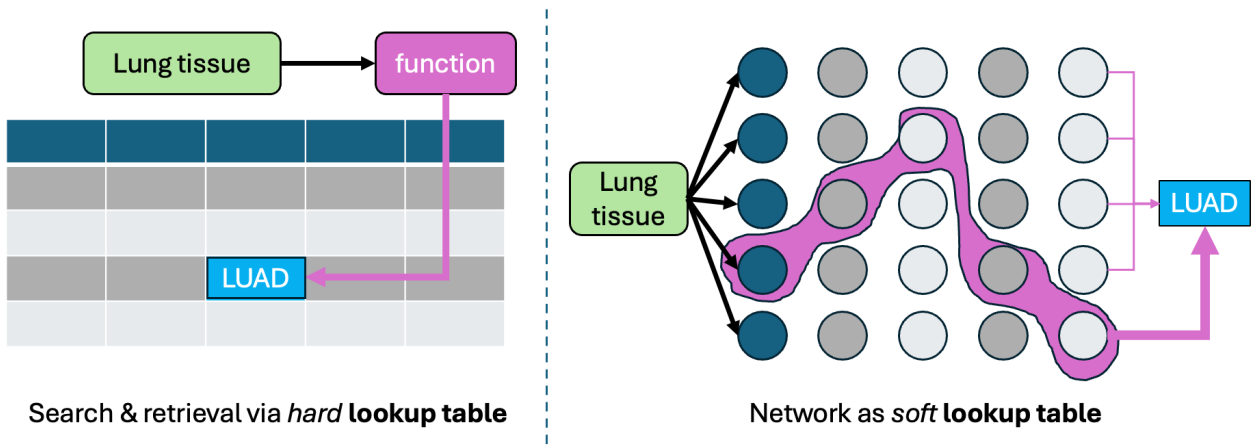

Figure 6. hard versus soft lookup tables for search. In conventional search a query "Lung tissue" is passed through 'function' (a mathematical function of some sort) to find the location of the right item 'LUAD' (lung adenocarcinoma) associate with "Lung tissue". A network can be understood as soft lookup table whereas the function is implemented in network itself and in a complex trajectory in the weight space of the network (concepts are oversimplified to illustrate the connection between explicit and implicit search).

A foundation model (FM), mostly based on transformer architectures, generally refers to a large language model (LLM) or a large vision-language models (LVLM) that serves as the basis for various downstream tasks concerning text and or images. Many specialized models can be derived from a foundation model. A FM is trained on an extremely large body of data, mostly using unsupervised learning (e.g., self-supervised learning). A FM captures patterns, relationships, and semantic representations within the data such that it can generate new data, which is the most fascinating (and most ethically sensitive) aspect of FMs. For instance, in contrast to previous language models, LLMs as text-specialized FMs can generate meaningful and context-aware responses, yielding suggestions that can be exploited for many downstream tasks such as summarization. FMs, such as OpenAI's chatGPT, have demonstrated significant advancements in the field of NLP and have enabled the development of various novel applications, chatbots, language assistants, and intelligent systems that can understand and generate human-like text. Other examples of FMs include GPT-3 (Generative Pre-trained Transformer 3), and BERT (Bidirectional Encoder Representations from Transformers). As for images, alone or in conjunction with text, one may mention FMs like CLIP (Contrastive Language-Image Pretraining) and SAM (Semantic Alignment through Multimodal Contrastive Learning).

The connection between image search and foundation models can manifest in two general scenarios: one involves extracting deep features using a foundation model and utilizing them to index images for search [Alfasly 2023, Sain2023]. Properly trained foundation models can offer expressive features, significantly enhancing search and retrieval capabilities. A second scenario involves the integration of search and deep models, where foundation models rely on Retrieval-Augmented Generation (RAG) [Lewis2020]. Among other applications, search and retrieval mechanisms can aid foundation models in tasks such as "source attribution" [Kamalloo2023] and fact-checking results by providing relevant information from small or large datasets, thereby reinforcing the foundation model's robustness. RAG has the potential to address the "black box" disadvantage inherent in deep models, a limitation that persists in various forms even for foundation models.

Recently, some researchers have collected online images (e.g., from Twitter and PubMed) to re-train CLIP for histopathology [Huang2023, Lu2023]. The general wisdom is that employing online images is not a suitable venue: *Gargabe in, Garage out*. Validations with high-quality clinical data have clearly shown these models will not provide any value in pathology [Alfasly2024]. When designing and training FMs for pathology with high-quality data, one has to redefine the requirements for image search. In this case, the tasks will be intrinsically multimodal and the Divide & Conquer tasks will need to be implemented within the pathology-aware FM. A new type of search will then perhaps be part of a FM-derived question and answering. The most crucial concern in this regard will be the "generative" aspect of FMs which is undesired in medicine when it degenerates into hallucinations. It's essential to note that foundation models with conversational capabilities are closely tied to search, serving as implicit information retrieval. Table 4 offers a comparison between search and foundation models.

Table 4 – Comparison of "information retrieval" versus "foundation models" as two technologies that may be used to assist pathologists in decision making (also see [Tizhoosh2024]).

|  | **Information Retrieval (IR)** | **Foundation Models (FMs)** |
|---|---|---|
| **Model Size** | Small | Very Large |
| **Size of Dataset Needed** | Small | Very Large |
| **Computational Footprint** | Small | Very Large |
| **Strength** | Convinces through retrieving evidence | Convince through knowledgeable conversations |
| **Disease Type Suitability** | All diseases including rare cases with only a few examples | Mainly common diseases with a lot of data available |
| **Information Processing Type** | Explicit Information Retrieval | Implicit Information Retrieval |
| **Source Attribution for responses to query** | Visible; Accessible; Explainable | Invisible; Not accessible; Not easily explainable |
| **Maintenance** | <ul><li>Low dependency on hardware updates</li><li>Cases can be added/deleted easily.</li><li>Re-indexing has moderate computational costs</li></ul> | <ul><li>High dependency on hardware updates</li><li>High efforts for prompting to customize for specific tasks</li><li>Expensive re-training cycles may be necessary</li></ul> |

## Summary and Conclusions

There has been minimal innovation in the area of "Divide" for efficient and reliable processing of WSIs. HBIR and SMILY lack this feature. Yottixel is the only search engine with a novel and unsupervised Divide, namely the mosaic. SISH and RetCCL rely on Yottixel. BoVW offers a different approach to Divide and deserves more attention. Clearly, there is an urgent need for new Divide strategies for patching WSIs that satisfy the following conditions: universality, biopsy independence, diagnostic inclusion, high-speed, and efficiency. Similarly, there has been little progress in the field of "encoding", which involves converting deep features into a more compact and less storage-hungry search engine like Yottixel's binary vectors (barcodes). It would be beneficial to develop a solution that eliminates the need for feature binarization altogether, although this task may prove more challenging than devising novel binarization methods not patented by industry, as is the case for Yottixel's barcodes. Multimodal search schemes are sorely lacking, highlighting a significant deficiency in the research community. While WSIs are valuable, their retrieval potential is limited as a single modality. The future lies in *searching more comprehensively for patient data*, and patient representation requires the integration of multiple modalities. Lastly, the pathology community should strive to develop foundation models based on large-scale, high-quality multimodal data. Such models would not replace the need for search but open new horizons for search capabilities in laboratory medicine and pathology.